\documentclass[%
jmp,%
 sd,%
 amsmath,amssymb,
 reprint,%
]{revtex4-1}

\usepackage{graphicx}
\usepackage{dcolumn}
\usepackage{bm}

\begin{document}

\preprint{AIP/123-QED}

\title[Brane Kantowski-Sachs Universe with Linear Equation of State and Future Singularity]{Brane Kantowski-Sachs Universe with Linear Equation of State and Future Singularity} 

\author{G S Khadekar}
 \email{E-mail: gkhadekar@yahoo.com}
\affiliation{ 
Department of Mathematics, Rashtrasant Tukadoji
Maharaj Nagpur University,\\
 Mahatma Jyotiba Phule Educational Campus, Amravati Road, Nagpur-440033 (INDIA)
}%


\date{\today}

\begin{abstract}
We consider the brane Kantowski-Sachs universe when bulk space is five dimensional anti-de Sitter obeying a linear equation of state of the form $p=(\gamma -1)\rho +p_{0}$, where $ \gamma,\:p_{0}$ are a parameters. In this framework we discuss the brane Kantowski-Sachs cosmology and compare it with standard Kantowski-Sachs cosmology in Einstein theory for some specific matter choice. For several specific choice of relation between energy and pressure, the behavior of the scale factors at early time is found. It is shown that  the brane Kantowski-Scahs cosmology may  quickly become isotropic for some choices of matter. We also discuss the universe fate of Kantowski-Sachs cosmological model by means of the rip or singularity in both the theories.
\end{abstract}

\pacs{Valid PACS appear here}
\maketitle

\section{\label{sec:level1}Introduction}

The brane world scenario assumes that our universe is a four dimensional space-time, a 3-brane, embedded in the five dimensional (5D) bulk space-time. Moreover inspired by the string theory/M-theory, the $Z_{2}$-symmetry with respect to the brane is imposed \cite{Horava}. Randall and Sundrum \cite{Randall}$^{-}$\cite{Randall1} have shown that a scenario with an infinite fifth dimension in the presence of brane can generate the theory of gravity which mimics purely four dimensional gravity, both with respect to the classical gravitational potential and with respect to gravitational relation. 
\par In the standard as well as in the brane cosmology it is quite possible that early universe could be the anisotropic one. The behavior of an anisotropic Bianchi type I brane-world in the presence of inflationary scalar fields has been considered by Maartens et al. \cite{Maarten}. Also Obukhov et al.\cite{Obukhov} considered the brane Kantowski-Sachs (KS) universe and compared the KS brane cosmology with KS cosmology for Einstein theory for several classical matter choices obeying the barotropic equation of state of the form $p=(\gamma-1)\rho$.
\par The cosmological observation indicate that the expansion of our universe accelerated [\cite{Totani}$^{-}$\cite {Riess}]. In the last decade lots of work on its possible mechanism, such as extended gravity \cite{Meng}, modifying equation of state are to explain the cosmic acceleration expansion observed. To over-come the drawback of hydrodynamics instability, Babichev et al. \cite{Babichev} was proposed a linear equation of state of more general from $p=\alpha(\rho -\rho_{0})$ and this form incorporated into cosmological model can describe the hydrodynamical stable dark energy behaviors. Meng et al.\cite{Meng} solved the Friedmann equations with both more general equation of state of the  form 
\begin{equation}
 \label{eq1} p=(\gamma -1)\rho +p_{0}, \;\;\;\;\; 1 \le \gamma \le 2,
\end{equation}
where $\gamma,\: p_{0}$ are a parameters, and the bulk viscosity, and discussed the acceleration expansion of the universe evolution and future singularity for the model.
\par In this paper we generalize the work of Obukhov et al. \cite{Obukhov} by assuming more general equation of state (EOS) of the form Eq.(1)  and compare  the brane Kantowski-Sachs (KS) cosmology with KS cosmology for Einstein theory for several specific choices of relation between energy and pressure and also discuss the fate of universe evolution by means of the rip or singularity.
\section{\label{sec:level2}Brane World and Kantowski-Scahs Geometry}
The brane world scenario was proposed by Randall and Sandrum \cite{Randall1}. They have shown that the in a five dimensional space-time (called bulk) it is possible to confine the matter field in four dimensional hypersurface called (3-brane).  The effective action in bulk is \cite{Maeda}  
$$S= \int d^5\sqrt-g_{5} \left(  \frac{1}{2k_{5}^2} R_{5}-\Lambda_{5} \right) $$
\begin{equation}
 \label{eq2} 
+ \int_{\chi=0} d^4\sqrt-g \left(  \frac{1}{2k_{5}^2} K^{\pm} +\lambda + {L}^{matter} \right). 
\end{equation}
Here any quantity in the bulk is marked with subscript $'5'$ with $k_{5}^2= 8\pi G_{5}$ being the five dimensional gravitational coupling constant. The coordinate on the bulk are  denoted by   $x^{\nu}, (\nu = 0,1,2,3, 4)$ while those on the brane as the $x^{\mu}, (\mu = 0,1,2,3)$. $R_{5}$ is the 5D intrinsic curvature in the bulk and $K^{\pm}$ is the intrinsic curvature on either side of the brane. 
\par On the  5D space-time (the bulk),  with the negative vacuum energy $\Lambda_{5}$ as the only source of the gravitational field the Einstein field equations are given by 
$$ ^{(5)}G_{IJ}= k_{5}^{2}, $$
\begin{equation}
\label{eq3}
^{(5)}T_{IJ}= - \Lambda_{5} + ^{(5)} g_{IJ} + \delta(\chi)[-\lambda g_{IJ} + T^{matter}_{IJ}].
\end{equation} 
In this space-time a brane is a fixed point of $Z_{2}$ symmetry.
\par Assuming a metric of the form $ds^{2}= (n_{I}n_{J}+ g_{IJ}) dx^Idx^J$, with $n_{I}dx^{I}=d\chi $ the unit normal to the $\chi=$ constant hypersurface and $g_{IJ}$ the induced metric on $\chi =$ constant hypersurfaces, the effective four dimensional gravitational equations on the brane takes the form \cite{Sasaki}$^{-}$\cite{Maartens}  
\begin{equation}
\label{eq4}
G_{\mu\nu}= -\Lambda g_{\mu\nu} + k_{4}^2 T_{\mu \nu} + k^{4}_{5} S_{\mu \nu}- E_{\mu\nu},
\end{equation}
where $G_{\mu\nu}$ and $T_{\mu\nu}$ are the Einstein and energy momentum tensors and $S_{\mu\nu}$ is the quadratic contribution on $T_{\mu\nu}$ defined as 
$$ S_{\mu\nu}=\frac{1}{12} T T_{\mu\nu} -\frac{1}{4} T^{\alpha}_{\mu} T_{\alpha\nu}+ \frac{1}{24}  g_{\mu\nu} ( 3 T^{\alpha\beta}T_{\alpha\beta}-T^2),$$ 
and $\Lambda= k^{2}_{5}(\Lambda_{5} + k^{2}_{5} \lambda^{2}/6), \; k^{2}_{4}=k^{2}_{5} \lambda^{2}/6 $ and $E_{IJ} = C_{IAJB}n^{A}n^{B}$. $C_{IAJB}$ is the five dimensional Weyl tensor in the bulk and $\lambda$ is the vacuum energy on the brane. $T_{\mu\nu}$ is the matter energy momentum tensor on the brane with the components $T^{0}_{0}=-\rho, T^{1}_{1}=T^{2}_{2} =T^{3}_{4}= p$ and $T=T^{\mu}_{\mu}$ is the trace of the energy momentum tensor. In this paper we restrict our analysis to a flat bulk geometry with $C_{IAJB}=0.$
\par We will consider  the brane metric in Kantowski-Sachs from \cite{Nojiri}$^{-}$\cite{Kantowski}
\begin{equation}
 \label{eq5} ds^{2}= -dt^{2}+R_{1}^{2}(t)dr^{2}+ R_{2}^{2}\left(d\theta ^{2} + sin^{2}\theta d\phi^{2}\right). 
\end{equation}
For later convenience, we will introduce the following variables
\begin{equation}
 \label{eq6}  V= R^3= R_{1}R_{2}^2,\;\; A_{m}=\Sigma_{i=1}^{i=3}\frac{(H_{i}-H)^2}{3H^2},
\end{equation}
\begin{equation}
 \label{eq7} H_{i}= \frac{\dot R_{i}}{R_{i}},\;\;\; H= \frac{1}{3}(H_{1}+2H_{2})=\frac{\dot V}{3V},\;\; q= \frac{d}{dt}(H^{-1})-1,
\end{equation}
where $R_{1},R_{2}$ are scale factors, $R$ average scale factor, $V$ is the volume scale factor, $H_{1}, H_{2}$ are Hubble factors, $H$ is average Hubble factor, $q$ is deceleration parameter, and $A_{m}$ anisotropy parameter.
\section{Conventional Einstein Theory with Linear Equation of State}
For the conventional Einstein theory (CET) the Einstein field equations and energy conservation law  
\begin{equation}
 \label{eq8}  G_{\mu\nu} =-\Lambda g_{\mu\nu} + k_{4}^{2}T_{\mu\nu},\;\;\; {\bigtriangledown}_{\mu}\: T^{\mu\nu}=0,
 \end{equation}
with $ G_{\mu\nu}$ being the Einstein tensor (4D), $\Lambda$ the cosmological constant and $k_{4}$ the gravitational coupling, $k^{2}_{4}= 8 \pi G.$
\par Einstein field equations and conservation law Eq.(8) for the Kantowski-Scahs universe Eq. (5) with  linear equation of state (LEOS) of  the form Eq. (1) becomes 
\begin{equation}
 \label{eq9}  \frac{\dot R_{2}^{2}}{R_{2}^{2}}+2 \frac{\dot R_{1}\dot R_{2}}{R_{1}R_{2}}+ \frac{1}{R_{2}^{2}}= \Lambda + k_{4}^{2}\rho,
 \end{equation}
\begin{equation}
 \label{eq10}  \frac{2\ddot R_{2}}{R_{2}}+ \frac{\dot R_{2}^{2}}{\dot R_{2}}+ \frac{1}{R_{2}^{2}}= \Lambda + k_{4}^{2}\rho (1-\gamma)-k_{4}^{2}p_{0} ,
 \end{equation}
\begin{equation}
 \label{eq11}  \frac{2\ddot R_{1}}{R_{1}}+\frac{\ddot R_{2}}{R_{2}}+ \frac{\dot R_{1}\dot R_{2}}{R_{1}R_{2}}= \Lambda + k_{4}^{2}\rho (1-\gamma)-k_{4}^{2}p_{0} ,
 \end{equation}
 \begin{equation}
 \label{eq12}  \dot\rho + (\gamma \rho +p_{0})\left(\frac{\dot R_{1}}{R_{1}} +2 \frac{\dot R_{2}}{R_{2}}\right).
 \end{equation}
  First of all, Eq. (12) can be easily solved to obtain the time evolution law of the energy density of the fluid
  \begin{equation}
 \label{eq13}  \rho = \rho_{0} V^{-\gamma}- \gamma^{-1} p_{0}, \;\;\; \rho_{0}= constant > 0.
 \end{equation}
 We can write the above field equations in the another form
 \begin{equation}
 \label{eq14}  \frac{d}{dt}\left({VH_{1}}\right)= \left[\Lambda -\frac{k_{4}^{2}p_{0}}{\gamma}\right]V +\frac{1}{2}k_{4}^{2}\:V^{(1-\gamma)}\:\rho_{0}(2-\gamma),
\end{equation}
 \begin{equation}
 \label{eq15}  \frac{d}{dt}\left({VH_{2}}\right)= \left[\Lambda -\frac{k_{4}^{2}p_{0}}{\gamma}\right]V +\frac{1}{2}k_{4}^{2}\:V^{(1-\gamma)}\:\rho_{0}(2-\gamma)-R_{1},
\end{equation}
 \begin{equation}
 \label{eq16}  3\dot H +H_{1}^2 +2H_{2}^2= \left[\Lambda -\frac{3k_{4}^{2}p_{0}}{2\gamma}\right] +\frac{1}{2}k_{4}^{2}\:V^{-\gamma}\:\rho_{0}(2-3\gamma).
\end{equation}
From Eq. (14) and Eq. (15), the equation for $V$ can be written as  
  \begin{equation}
 \label{eq17} \ddot V = 3\left[\Lambda -\frac{k_{4}^{2}p_{0}}{\gamma}\right]V +\frac{3}{2}k_{4}^{2}\:V^{(1-\gamma)}\:\rho_{0}(2-\gamma)-2R_{1}.
\end{equation}
 After first integration 
 \begin{equation}
 \label{eq18} \dot V^2 = 3\left[\Lambda -\frac{k_{4}^{2}p_{0}}{\gamma}\right]V^2 +3 k_{4}^{2}\:V^{(2-\gamma)}\:\rho_{0}-4\int R_{1}dV.
\end{equation}
Again from Eq. (14) and Eq. (15)
  \begin{equation}
 \label{eq19} \frac{d}{dt}\left[{V(H_{1}- H_{2})}\right]= -R_{1}.
\end{equation}
 It follows that 
  \begin{equation}
 \label{eq20} 
 H_{1} = H+ \frac{2}{3V}K, \;\; H_{2} = H- \frac{1}{3V}K, \;\; K=\int R_{1}dt.
\end{equation}
\section{Asymptotic Behavior in CET}
In this section we consider the asymptotic behavior for $V$ in CET.
\subsection{ For large $V$ }
If $V$ is large then RHS of Eq. (18) $\rightarrow 3\left[\Lambda -\frac{k_{4}^{2}p_{0}}{\gamma}\right]V^2 $, when $ \Lambda -\frac{k_{4}^{2}p_{0}}{\gamma} >0. $ This leads to the volume scale factor of our universe to expand, exponentially in the large time limit i.e. $V \propto exp\left[\sqrt{3\left(\Lambda -\frac{k_{4}^{2}p_{0}}{\gamma}\right)}t\right]$ and $ H_{1} = H_{2}$.
\par By using Eq. (6) and (7) we get 
 \begin{equation}
 \label{eq21} 
 R_{1} = \frac{a}{b^2}exp[\sqrt{\left(\Lambda_{0}/3 \right)}t],\;\;\; R_{2} = b \;exp[\sqrt{\left(\Lambda_{0}/3 \right)}t],
\end{equation}
where $a, b$ are constants and $\Lambda_{0}= (\Lambda -k_{4}^{2}p_{0}\gamma^{-1}).$
\par Therefore, in this limit the mean  anisotropic parameter $A_{m}$ decays to zero exponentially and the deceleration parameter $q$ becomes negative i.e. $q<0.$ Hence the universe can be isotropized dynamically and undergoes an accelerated expansion in the large time limit due to the presence of the term $\Lambda_{0} > 0$ when $p_{0}< 0$. In fact, the value of $\Lambda_{0}$ can only change the expanding rate of the universe but not affect its effect of isotropization in general. The behavior of the system does not depends on $\gamma$ and the anisotropic CET universe becomes isotropic one for large $V$ due to the classical  matter effects.  
\subsection{ For extremely  small $V$ }
In this case the properties of the CET universe will depends on the value of $\gamma.$ When $V$ is extremely small at the limit $t \rightarrow 0,$ the  RHS of Eq. (18) $\rightarrow 3 k_{4}^{2} \rho_{0}V^{(2-\gamma)} + b, $ where $b$ is constant.  Therefore, the mean anisotropy parameter $A_{m}$ at the early universe tends to zero.
 
\par Hence from Eq. (18)
  \begin{equation}
 \label{eq22} 
  \dot V = \sqrt{ 6 a \rho_{0}V^{(2-\gamma)} + b },
\end{equation}
where $a=\frac{1}{2}k_{4}^{2} \rho_{0}.$ 
\par For $\gamma =3/2,$ the solution takes the form 
 
  $$V = \frac{9}{4}a^{3/2}(t-t_{0})^{4/3}, \;\;\; R_{1}= \frac{9}{4c^2} a^{3/2}(t-t_{0})^{4/9},$$
   \begin{equation}
 \label{eq23} 
   R_{2}= c(t-t_{0})^{4/9},
\end{equation}
where $c$ is the constant of integration and also supposed that $V(t_{0})= 0.$
\par For $\gamma =1,$ the solutions takes the form 
 \begin{equation}
 \label{eq24} 
  V \propto (t-t_{0})^2, \;\;\; R_{1} \propto (t-t_{0})^{2/3},\;\; R_{2}\propto (t-t_{0})^{2/3}.
\end{equation}
In this case the solution is also isotropic. Hence it is observed that for the large and the sufficiently small  $V$ the solution becomes isotropic. It is note that even for small cosmological time the process of isotrozation starts quickly \cite{Obukhov}. 
\section{Brane Kantowski-Sachs Universe with Linear Equation of state}
In this section we consider the brane cosmology with zero Weyl tensor i.e. $ C_{IAJB} n^{A} n^{B}=0.$
\par The gravitational field equations (4) and the conservation law on the brane with LEOS (1) takes the form
\begin{equation}
 \label{eq25}  \frac{\dot R_{2}^{2}}{R_{2}^{2}}+2 \frac{\dot R_{1}\dot R_{2}}{R_{1}a_{2}}+ \frac{1}{R_{2}^{2}}= \Lambda + k_{4}^{2}\rho + \frac{1}{12}k_{5}^{4}\rho^2 ,
 \end{equation}
\begin{equation}
 \label{eq26}  \frac{2\ddot R_{2}}{R_{2}}+ \frac{\dot R_{2}^{2}}{\dot R_{2}}+ \frac{1}{R_{2}^{2}}= (\Lambda-k_{4}^{2}p_{0}) - k_{4}^{2}\gamma \rho +\frac{1}{12}k_{5}^{4}\rho^2(1-2\gamma),
 \end{equation}
\begin{equation}
 \label{eq27}  \frac{2\ddot R_{1}}{R_{1}}+\frac{\ddot R_{2}}{R_{2}}+ \frac{\dot R_{1}\dot R_{2}}{R_{1}R_{2}}= (\Lambda-k_{4}^{2}p_{0}) - k_{4}^{2}\gamma \rho +\frac{1}{12}k_{5}^{4}\rho^2(1-2\gamma),
 \end{equation}
 \begin{equation}
 \label{eq28}  \dot\rho + (\gamma \rho +p_{0})\left(\frac{\dot R_{1}}{R_{1}} +2 \frac{\dot R_{2}}{R_{2}}\right)=0.
 \end{equation}
Here we consider  $ \left(\frac{k_{5}^4 p_{0}}{6}\right)= k_{4}^{2}$ and $p_{0}$ plays the role of $\lambda,$ the vacuum energy on the brane.
\par From Eq. (28) after integrating with the help of Eqs. (6) and (7) we get
 \begin{equation}
 \label{eq29}  \rho = \rho_{0} V^{-\gamma}- \gamma^{-1} p_{0}, \;\;\; \rho_{0}= constant > 0.
 \end{equation}
The above field equations in another form as 
  $$ \frac{d}{dt}\left({VH_{1}}\right)= \left[\Lambda -\frac{k_{4}^{2}p_{0}(2\gamma-1)}{2\gamma^2}\right]V $$
   \begin{equation}
 \label{eq30}
  +\frac{1}{2\gamma}k_{4}^{2}\:V^{(1-\gamma)}\:\rho_{0}(1-\gamma)(\gamma-2) + \frac{k_{5}^{4}}{12}\rho_{0}^2 (1-\gamma)V^{(1-2\gamma)},
\end{equation}
  $$\frac{d}{dt}\left({VH_{2}}\right)= \left[\Lambda -\frac{k_{4}^{2}p_{0}(2\gamma-1)}{2\gamma^2}\right]V $$
  \begin{equation}
 \label{eq31} +\frac{1}{2\gamma}k_{4}^{2}\:V^{(1-\gamma)}\:\rho_{0}(1-\gamma)(\gamma-2) + \frac{k_{5}^{4}}{12}\rho_{0}^2 (1-\gamma)V^{(1-2\gamma)}-R_{1},
\end{equation}
 $$  3\dot H +H_{1}^2 +2H_{2}^2= \left[\Lambda -\frac{k_{4}^{2}p_{0}(2\gamma^2 +2\gamma-1)}{2\gamma^2}\right] $$
 \begin{equation}
 \label{eq32} + k_{4}^{2}\:V^{-\gamma}\:\rho_{0}(1+3\gamma)(1-3\gamma)+\frac{1}{12}k_{4}^{2}\:V^{-2\gamma}\:\rho_{0}^2(1-3\gamma).
\end{equation}
From Eqs. (30) and (31) the equation for $V$ can be obtained as 
$$ \ddot V = 3\left[\Lambda -\frac{k_{4}^{2}p_{0}(2\gamma^2+2\gamma-1)}{2\gamma^2}\right]V $$
\begin{equation}
 \label{eq33} +\frac{3}{2\gamma}k_{4}^{2}\:V^{(1-\gamma)}\:\rho_{0}(1-\gamma)(\gamma-2)-2R_{1}+\frac{1}{4}k_{4}^{2}\:V^{(1-2\gamma)}\:\rho_{0}^2(1-\gamma).
\end{equation}
After first integration we get
 \begin{equation}
 \label{eq34} \dot V = \sqrt{F(V)},
 \end{equation}
 where
 $$ F(V) = 3\left[\Lambda -\frac{k_{4}^{2}p_{0}(2\gamma-1)}{2\gamma^2}\right]V^2- 3k_{4}^{2}\:V^{(2-\gamma)}\:\rho_{0}\left(\frac{1-\gamma}{\gamma}\right)$$
 \begin{equation}
 \label{eq35}+\frac{1}{4}k_{4}^{2}\:V^{2(1-\gamma)}\:\rho_{0}^2-4\int R_{1}dV.
\end{equation}
Again from Eqs. (30) and (31)
\begin{equation}
 \label{eq36} 
 H_{1} = H+ \frac{2}{3V}K, \;\; H_{2} = H- \frac{1}{3V}K, \;\; K=\int R_{1}dt.
\end{equation}
\section{Asymptotic Behavior in Brane Cosmology}
In the following part we will discuss the asymptotic behavior of for large and small  $V. $ 
\par In the brane cosmology the asymptotic behavior for the large $V$ is similar to the CET. However, for the extremely small $V$ one obtains difference in comparison with CET.
\subsection{For extremely small  $V$}
 For extremely small $V$ and $\gamma=1$ the RHS of Eq. (34) $\rightarrow (a^2 +b),$ where $a^2=\left(\frac{k_{5}^{4}\:\rho_{0}^2}{4}\right)$ and $b$ is constant.
\par Then from Eq. (34) 
\begin{equation}
 \label{eq37} 
 V = C_{1} +\sqrt{a^2 +b}\;t,
\end{equation}
where $C_{1}$ is the constant of integration. 
\par From Eq. (36)
\begin{equation}
 \label{eq38} 
 R_{1} = C_{1}V^{1/3}exp\left[ \frac{2K}{3}\int V^{-1}F(V)^{-1/2}dV\right].
\end{equation}
After substituting the value of $V$ from Eq. (37) into Eq. (38) we get
\begin{equation}
 \label{eq39} 
 R_{1} = C_{2}[C_{1} +\sqrt{a^2 +b}\;t]^{\frac{1}{3} +\frac{2K}{3\sqrt{a^2 +b}}},
\end{equation}
\begin{equation}
 \label{eq40} 
 R_{2} = \frac{1}{\sqrt{C_{2}}}[C_{1} +\sqrt{a^2 +b}\;t]^{\frac{1}{3} -\frac{K}{3\sqrt{a^2 +b}}}.
\end{equation}
For $\gamma =3/2$, we get 
\begin{equation}
 \label{eq41} 
 R_{1} = \frac{1}{C_{2}^2} \left(\frac{3a}{2} \right)^{2/3}(t-t_{0})^{2/9},
\end{equation}
\begin{equation}
 \label{eq42} 
 R_{2} = C_{2}(t-t_{0})^{2/9},\;\;V = \left(\frac{3a}{2}\right)^{2/3}(t-t_{0})^{2/3}.
\end{equation}
Here $C_{1}, C_{2}$ are constants. 
\section{Conclusions and Discussions}
In the present work we have compared the KS cosmological model in CET and the brane theory by using LEOS. The two different theories give, however completely different anisotropic expansion at the very early stags of evolution. 
From Eq. (4), it is easy to realize that the brane world scenario is different from the CET by two parts: (i) the matter fields contribute local ''quadratic'' energy momentum tensor correction via the tensor $S_{\mu\nu}$, and (ii) there are ``nonlocal'' effects from the free field in the bulk, transmitted via projection of the bulk Weyl tensor $E_{\mu\nu}.$ Therefore, the CET can be treated as a limit of the brane theory by taking $E_{\mu\nu}=0$ and $k_{5} \rightarrow 0$  with properly adjusted values of the constant $k_{4}$ and $\Lambda$ \cite{Chen}. The CET can be reproduced by imposing the limit $k_{5} \rightarrow 0.$
\par Indeed, for the CET the anisotropic expansion tends to be large at the very early stage. In the another words, the universe has to begin from highly anisotropic initial expansion and then decay to zero as the time increase. In the other hand, in the brane cosmology, due to the quadratic correction which sufficiently changes the early time behavior of the universe. 
\par It is observed that for $\gamma=3/2,$ brane KS  universe become isotropic at early time analogous to CET case. The nature of the scale factors for $\gamma=3/2,$ in both the cases are slightly different, which shows the role of five dimensional bulk. For $\gamma=1,$ KS cosmology for CET become isotropic and brane KS universe remains anisotropic. It is observed that in both the model we don't get the exact solution. However, when $R_{1}$ = constant, $\Lambda=0,\; p_{0}=0$ and   $\gamma=2,$ we get the exact solution only in case of  CET. Therefore, the equation of state most appropriate to describe the high density regime of the early universe is the stiff Zeldovich one, with  $\gamma=2.$  \cite{Chen1}.
\par It is worth noting that for $p=p_{0}$ i. e. $\gamma =1,$ mean anisotropy of the KS brane model behave similar to the models CET where the mean anisotropy parameter $A_{m}$ is large in the very early time.
\par We are able to investigate the fate of our cosmological models by means of the rip or singularity. The singularity behavior is classified in \cite{Nojiri1} $^{-}$\cite{Khadekar}: 
\begin{itemize}
	\item Type I (Big Rip): For $ t \rightarrow t_{s},\; a \rightarrow \infty,\;  \rho \rightarrow \infty,$ and $|p| \rightarrow \infty, $
	\item Type II (Sudden Singularity): For $ t \rightarrow t_{s}, \; a \rightarrow a_{s} , \; \rho \rightarrow \rho_{s},$ and $|p| \rightarrow \infty,  $
	\item Type III: For $ t \rightarrow t_{s}, \; a \rightarrow a_{s} , \; \rho \rightarrow \infty$ and $|p| \rightarrow \infty,$  
	\item Type IV: For $ t \rightarrow t_{s},\;  a \rightarrow a_{s} , \; \rho \rightarrow 0,$ and $|p| \rightarrow 0 $ and $p$ and $\rho$ are finite and higher order derivatives of $H$ diverge. 
\end{itemize}
For CET with $\gamma=3/2$ and  $\gamma=1$ at $t\rightarrow t_{0}$, the average scale factor $R \rightarrow 0$ and $\rho \rightarrow \infty$, so the type III singularity occurs. Also  for the KS brane cosmology with $\gamma=3/2,$ there will be again a type III singularity. However, for $\gamma=1$ in KS brane  cosmology there is no future singularity. This shows the qualitative difference between CET and brane cosmology.


\nocite{*}

\begin{thebibliography}{00}  
\bibitem{Horava}P. Horava E. Witten, {\it Nucl.Phys.} {\bf B460}  506 (1996). 
\bibitem{Randall}L. Randall and R Sundrum, {\it Phys Rev. Lett.} {\bf 83}  3370 (1999).
\bibitem{Randall1}L. Randall and R. Sundrum, {\it Phys. Rev. Lett.} {\bf 83}  4690 (1999).
\bibitem{Maarten}R.  Maartens, V. Sahni, and T. D. Saini, {\it Phys. Rev.} {\bf D 63}  063509 (2001).
\bibitem{Obukhov}V. V. Obukhov, A. V. Makarenko and K. E. Osetrin, {\it J. Phys A:  Math. Gen.} {\bf 39}  6635 (2006).
\bibitem{Perlmutter}S. Perlmutter et at., {\it Nature (London)} {\bf 391}  51 (1998).
\bibitem{Totani}T. Totani, Y. Yoshii and K. Sato, {\it Astrophys J.} {\bf 483} L75 (1999) .
\bibitem{Riess}A. G. Riess et al., {\it Astrophys J.}  {\bf 116}  1009 (1998).
\bibitem{Meng}X. H. Meng, J. Ren  and M. G. Hu, {\it Commun. Theor. Phys.} {\bf 47}  379 (2007).
\bibitem{Babichev}E. Babichev, V. Dokuchaev and Y. Eroshenko, {\it Class. Quantum Grav.} {\bf 22} 13  (2005).
\bibitem{Maeda}K. Maeda and D. Wands, {\it Phys. Rev.} {\bf D 62} 124009 (2000).
\bibitem{Sasaki}M. Sasaki, T. Shiromizu and K. Maeda, {\it Phys. Rev.} {\bf D 62}  024008 (2000).
\bibitem{Shiromizu}T. Shiromizu, K. Maeda and M. Sasaki, {\it Phys. Rev.} {\bf D 62}  024012 (2000).
\bibitem{Maartens}R. Maartens, {\it Phys. Rev.} {\bf D 62}  084023 (2000).
\bibitem{Nojiri}S. Nojiri, O. Obergon, S. D. Odintos and K. E. Osetain, {\it Phys. Rev.} {\bf D 60}  024008 (1999).
\bibitem{Kantowski} R. Kantowski and R.  Sachs, {\it J. Math. Phys} {\bf 7}  443 (1966).
\bibitem{Chen}C. M. Chen and W. F. Kao, {\it hep-th/0201188} (2002).
\bibitem{Chen1}C. M. Chen, T. Harko  and M. K. Mak, {\it Phys. Rev. D} {\bf 64}  (044013) (2001).
\bibitem{Nojiri1}S. Nojiri, S. D. Odintsov and S. Tsujikawa, {\it Phys. Rev. D} {\bf 71} (063004) (2005).
\bibitem{Khadekar}G. S. Khadekar and N. V. Gharad, {\it The Open Astronomy Journal} {\bf 7}  1 (2014). 
\end{thebibliography}

\end{document}